\documentclass[conference]{IEEEtran}
\usepackage{wrapfig}
\usepackage{graphicx}
\usepackage{amsmath}
\usepackage{amssymb}
\usepackage{multirow}
\usepackage{pifont}
\usepackage{textcomp}
\usepackage{algorithmic}
\usepackage{color}
\usepackage{centernot}
\usepackage{verbatim}

\usepackage{enumitem}
\newtheorem{proposition}{Proposition}

\DeclareMathOperator*{\argmax}{arg\,max}
\DeclareMathOperator*{\argmin}{arg\,min}

\newcommand\T{\rule{0pt}{2.1ex}}
\newcommand\B{\rule[-0.7ex]{0pt}{0pt}}

\newcommand{\norm}[1]{\left\lVert#1\right\rVert}

\begin{document}
%
%\setcopyright{acmcopyright}

\title{An Integrated Framework for Competitive Multi-channel Marketing of Multi-featured Products
%Competitive Influence Maximization in Social Networks in the Presence of Other Advertising Channels
%\titlenote{Post acceptance of this paper, source code will be shared on \texttt{lcm.csa.iisc.ernet.in} for community contribution.}
}

%\numberofauthors{2} 
\author{
\IEEEauthorblockN{Swapnil Dhamal}
\IEEEauthorblockA{INRIA Sophia Antipolis M\'editerran\'ee, France\\
Email: swapnil.dhamal@inria.fr}
%\alignauthor
%\hspace{-0cm} Swapnil Dhamal \\%\titlenote{}\\
%%       \affaddr{Computer Science and Automation}\\
%       \affaddr{\hspace{-0cm} Indian Institute of Science}\\
%       \affaddr{\hspace{-0cm} Bangalore 560012, India}\\
%       \email{\hspace{-0cm} swapnil.dhamal@csa.iisc.ernet.in}
}
%%\date{30 July 1999}

\maketitle

\begin{abstract}
For any company, multiple channels are available for reaching a population in order to market its products. Some of the most well-known channels are (a) mass media advertisement, (b) recommendations using social advertisement, and (c) viral marketing using social networks. The company would want to maximize its reach while also accounting for simultaneous marketing of competing products, where the product marketings may not be independent. In this direction, we propose and analyze a multi-featured generalization of the classical linear threshold model. We hence develop a framework for integrating the considered marketing channels into the social network, and an approach for allocating budget among these channels.
\end{abstract}
\begin{IEEEkeywords}
Social networks, viral marketing, product features, mass media, social advertisement, budget allocation
\end{IEEEkeywords}
%}

% A category with the (minimum) three required fields
%\category{H.4}{Information Systems Applications}{Miscellaneous}
%A category including the fourth, optional field follows...
%\category{D.2.8}{Software Engineering}{Metrics}[complexity measures, performance measures]

%\terms{Theory}

%\keywords{Social networks, social advertising, mass media advertising, campaign, information diffusion, competitive influence maximization.}

%\vspace{-.5mm}
\section{Introduction
%\vspace{-.3mm}
}
\label{sec:intro}

Companies can market their respective products through several possible channels,
%There are several ways in which a company or a political party can advertise its products or ideologies, 
the most prominent being mass media advertising (television, radio, newspapers, etc.), fixed Internet banner ads, banner ads based on browsing history,
recommendations based on attributes 
%of a potential customer 
(location, age group, field of work, etc.), recommendations based on friends' purchases, 
%purchasing behaviors of the individual and its friends, 
 sponsored search ads, 
 %social advertising, 
 social media, 
%direct recommendations to friends of a node who has bought the product, 
sponsored Internet reviews, and so on. Potential customers or nodes also get indirectly influenced through their friends owing to word-of-mouth marketing.
% and reviews on the Internet. 
%Some of these ways are node-specific (targeted  based on node's profile and browsing behavior)
% (search engine ads, social ads, etc.), 
%while others are independent of them.
%(television ads, fixed Internet banner ads, etc.).
%
 In order to make optimal use of these channels, a company would  want to make the decision of how to invest in each channel, based on the investment strategy of competitors who also market their products simultaneously.

%2nd

%\subsection{Preliminaries}
%\label{sec:prelim}
This paper aims to present a framework for competitive influence maximization
%studying optimal strategies 
%(selecting seed nodes) for triggering viral marketing, 
in the presence of several marketing channels.
%, so as to maximize the spread of influence in the social network.
We focus on modeling three 
%types of marketing 
channels, namely, viral marketing, mass media advertisement, and recommendations  based on friends' purchases using social advertisement.
% in the presence of two other types of advertising channels, which are not explicitly based on the online node profile or browsing behavior:

\subsubsection{Viral Marketing}

In our context, a social network can be represented as a weighted, directed graph, consisting of nodes which are potential customers.
%Several models  have been proposed in the literature for studying information diffusion in such a social network \cite{networkscrowdsmarkets}. 
%
%The linear threshold (LT) model and the independent cascade (IC) model are two of the most extensively studied models for information diffusion in social networks.
The model we propose for influence diffusion in social network is a generalization of the well-studied linear threshold (LT) model
%, one of the most extensively studied models for information diffusion in social networks 
\cite{networkscrowdsmarkets}.
% \cite{granovetter1978threshold}. 
%
%\subsubsection{The Linear Threshold model}
\begin{comment}
In LT model, every directed edge $(v,u)$ has weight $b_{u,v} \geq 0$, which is the degree of influence that node $v$ has on node $u$, and every node $u$ has an influence threshold $\chi_u$. The weights $b_{u,v}$ are such that $\sum_v b_{u,v} \leq 1$. Owing to lack of knowledge about the thresholds, which are held privately by the nodes, it is assumed that the thresholds are chosen uniformly at random from $[0,1]$. The diffusion process starts at time step 0 with the initially activated set of seed nodes, and proceeds in discrete time steps, one at a time. In each time step, a node is influenced or activated if and only if the sum of influence degrees of the edges incoming from activated neighbors (irrespective of the time of activation of the neighbors) crosses its own influence threshold, that is, when
%
$
\sum_v b_{u,v} \geq \chi_u
$.
%
Nodes, once activated, remain activated for the rest of the diffusion process. 
% In any given time step, the recently activated nodes along with previously activated ones contribute to the diffusion process. 
The diffusion process stops when it is not possible to activate or influence any further nodes.
\end{comment}
%
Given such a  model, a company would want to select a certain number of seed nodes to trigger viral marketing so that maximum number of nodes get influenced (buy the product)
%. This problem is generally termed as {\em influence maximization} 
\cite{kempe2003maximizing}.

%\subsection{Motivation}
%\label{sec:motiv}

\begin{comment} %2nd
In this paper, we assume that nodes are anonymous, given the network, that is, if we interchange any two nodes, the dynamics of the diffusion remains unchanged. In other words, we ignore any node-specific information such as online node profile and browsing behavior. 
We focus on viral marketing in the presence of two other types of advertising channels, which are not explicitly based on the online node profile or browsing behavior:
\end{comment} %2nd

\subsubsection{Mass media advertisement}
This is one of the most traditional way of marketing where a company advertises its product to the masses using  well-accessible media such as television, radio, and newspaper. 
\begin{comment} %2nd
Even though the advertisement could be targeted towards a particular audience (like people watching a particular program on television or reading a particular page in newspaper), mass media generally allows potentially anyone to access the advertisement.
\end{comment} %2nd
The timing of when to show the ads is critical to ensure optimal visibility and throughput.

%In this paper, we make a simplifying assumption that all nodes  have an equal probability of getting exposed to mass media advertisements at any given time. 

\subsubsection{Social advertisement  based on friends' purchases}
While making purchasing decisions, nodes rely not only on their own preferences but also on their friends', owing to social correlation  due to homophily (bias in friendships towards similar individuals) \cite{chua2012generative}. This, in effect, can be harnessed to suggest products to a node based on its friends' purchasing behaviors.
%
%It has been empirically observed that homophily is one of the most prominent phenomena in social networks, that is, in a social network, there is a bias in friendships towards similar individuals. 
If a node has  high influence on its friend (which is accounted for in diffusion models like LT), it is likely that the two nodes are similar. However, if  the influence is low, it is not conclusive whether the  nodes are dissimilar. So in addition to the influence parameter considered in LT-like models, marketing in practice requires
%$b_{uv}$, let $h_{uv}$ (or $h_{vu}$) be 
another parameter that quantifies  similarity between nodes. 
%This parameter can be deduced using the likes and interests of the two nodes with respect to posts, pages, etc. on online social networking sites.
%
Note that since diffusion models do not consider this similarity, they alone cannot justify why two nodes having negligible influence on each other, display similar  behaviors. 
It has also been observed in Twitter that almost 30\% of information is attributed to factors other than network diffusion
%%
%It has also been observed that about 71\% of the information volume in Twitter can be attributed to network diffusion, while the remaining 29\% is due to factors external to the network 
\cite{myers2012information}.
The effect of such factors could hence be captured using the similarity parameter.
\vspace{-1.1mm}
\subsection{Related Work
\vspace{-1mm}
}
\label{sec:relevant}

%Dhamal and Narahari
%\cite{dhamal2013scalable} show how the ability of agents to aggregate information from their local social neighborhood can be used for efficiently aggregating the preferences of all the agents in a social network. 

%\cite{manchanda2008role}
%\cite{hauser2008defensive}
%\cite{saravanakumar2012social}
%\cite{trusov2009effects}
%\cite{chintagunta2010effects}
%\cite{peres2010innovation}
%\cite{ho2012customer}
%\cite{sonnier2011dynamic}
%\cite{easingwood1983nonuniform}
%\cite{kuksov2013advertising}

The problem of influence maximization is well-studied in  literature on social network analysis. 
%The two basic models of information diffusion are linear threshold model \cite{granovetter1978threshold} and independent cascade model \cite{goldenberg2001talk}. 
It is known that computing the exact value of the objective function for a given seed set (the expected number of influenced nodes at the end of  diffusion  that was triggered at that set), is \#P-hard under
% the IC model \cite{chen2010scalable} as well as 
the LT model \cite{chen2010scalablelt}.
%Chen, Yuan and Zhang \cite{chen2010scalable} show that computing exact influence $\sigma(S)$, given the initial set of influencers $S$, in general networks in the linear threshold model is \#P-hard.
However, the value can be well approximated  using sufficiently large number of Monte-Carlo simulations.
Though the influence maximization problem under   LT model is NP-hard,
%Kempe, Kleinberg, and Tardos 
%\cite{kempe2003maximizing} show NP-hardness of the influence maximization problem under the  LT 
%%and the IC
% model. The authors prove that 
  the  objective function is non-negative, monotone, and submodular; so greedy hill-climbing algorithm provides an
 approximation guarantee for its maximization
 % $(1-\frac{1}{e}-\epsilon)$-approximation algorithm for maximizing the same, where $\epsilon$ is very small for sufficiently large number of Monte-Carlo simulations 
 \cite{kempe2003maximizing}.
There exist  generalizations of LT model, e.g., general threshold model \cite{kempe2003maximizing}, extensions to account for time \cite{chen2012time}, and extensions to account for competition
% and decreasing cascade model \cite{kempe2005influential}.
%
%Borodin, Filmus, and Oren 
%\cite{borodin2010threshold} provide 
%Several natural extensions to the LT model have been proposed 
\cite{borodin2010threshold,pathak2010generalized}.
%and show that 
%%the  greedy approach cannot be used; they show that 
%%for a broad family of competitive influence models, it is NP-hard to achieve an approximation that is better than a square root of the optimal solution.
%% and hence suggest a natural model that is amenable to the greedy approach.
%
%Chen, Lu, and Zhang 
%\cite{chen2012time} 
%The problem of maximizing influence spread 
%in a social network 
%within a given deadline has also been studied, by extending the standard LT model to account for time \cite{chen2012time}.
% and IC models to account for time.
%
%There have been efforts to solve the problem efficiently by reducing the running time of greedy algorithm and proposing heuristics
%%Chen, Wang, and Yang 
%\cite{chen2009efficient}.
%improve the greedy algorithm by reducing its running time and also propose heuristics that improves influence spread as compared to classic degree and centrality-based heuristics. 
%
%
State-of-the-art heuristics such as
LDAG \cite{chen2010scalablelt} and Simpath \cite{goyal2011simpath}
perform  close to greedy algorithm while running several orders of magnitude faster.
%
%Chen, Yuan and Zhang \cite{chen2010scalable} show that computing exact influence, that is, the objective function, in general networks in the linear threshold model is \#P-hard and also propose a scalable influence maximization algorithm.
%
There also exist algorithms 
%Suri and Narahari 
%\cite{narayanam2010shapley} provide an algorithm 
%for influence maximization 
that provide good performance irrespective of the  objective function being submodular
%, unlike the greedy hill-climbing algorithm 
\cite{narayanam2010shapley}. 

The problem of competitive influence maximization wherein multiple companies market competing products using viral marketing has also been studied
%Bharathi, Kempe, and Salek 
\cite{bharathi2007competitive,goyal2012competitive}.
%study the problem of competitive influence maximization wherein multiple companies market competing products using viral marketing.
%; they provide an approximation algorithm for computing the best response to the strategy of competitors.
%
%
%Pathak, Banerjee, and Srivastava 
%\cite{pathak2010generalized} provide a generalized version of the LT model for multiple cascades on a network while allowing nodes to switch between them.
% wherein, the steady state distribution of a Markov chain is used to estimate highly likely states of the cascades' spread in the network.
%
%Myers and Leskovec 
%\cite{myers2012clash} develop a 
Also, more realistic models have  been developed, where influences not only diffuse simultaneously but  also interact
%, that is, compete or cooperate 
with each other 
%as they spread over the network
\cite{myers2012clash,zarezade2017correlated}. 

The impact of recommendations and word-of-mouth marketing on product sales revenue is  well studied in marketing literature
\cite{godes2012strategic,van2010viral,aral2011creating}. 
Biases in product valuation and usage decisions when agents consider a product that offers new features of uncertain value, have been investigated  \cite{meyer2008biases}.
%; agents display a high willingness to pay for new features.
%
%The sensitivity of diffusion dynamics with respect to uncertainty in the network structure has been studied in \cite{adiga2013sensitivity}.
%
%The problem of influence maximization has also been studied in the case where the network is either unavailable or insufficient to explain the underlying information diffusion phenomena \cite{wang2013sparse}.
%
%It has been observed that in making adoption or purchasing decisions, nodes rely not only on their own preferences, but also on their friends' owing to social correlation primarily caused by homophily \cite{chua2012generative}. This, in effect, can be used to suggest products to a node based on its and its friends' adoption or purchasing behaviors.
%Braun and Bonfrer 
It has also been discussed 
how marketers can apply latent similarities of customers for segmentation and targeting 
%activities
\cite{braun2011scalable}.
%develop a probabilistic framework for modeling agent interactions and find that the framework draws insights into latent similarities of customers; they also discuss how marketers can apply these insights to segmentation and targeting activities.
%
%There have been studies considering both geographical and social influence for point of interest recommendations in location-based social networks \cite{cheng2012fused}.
%
%

\begin{comment} %2nd
%Myers, Zhu, and Leskovec 
\cite{myers2012information} develop a model in which information can reach a node either via links of the social network or through external sources and find that factors external to the network play a significant role in information diffusion.
%significant  only about 71\% of the information volume in Twitter can be attributed to diffusion using the network, while 29\% is due to factors external to the network.
\end{comment} %2nd
 
%\cite{verhoef2015multi}

%\cite{jain2009self}
%\cite{vikander2012advertising}
%\cite{balachander2010bundle}
%\cite{berger2010positive}
%\cite{goldfarb2009measuring}
%\cite{luo2009quantifying}
%\cite{freimer2008try}
%The authors then show that social hubs significantly accelerate the diffusion process. They further distinguish between innovator and follower hubs and show that the former influence mainly the speed of adoption in a network, while the latter influence mainly the number of people that eventually adopt the innovation \cite{goldenberg2009role}.

\begin{comment}
\subsection{Our Contributions}
\label{sec:contrib}

\begin{itemize}
\item A vectors-based model that considers both personal aggregation and clash of contagions of multi-feature products
\item Integration of multiple channels into the social network
\item Cross entropy method
\end{itemize}
\end{comment}

%To the best of our knowledge, this work would be the first to study the problem of competitive influence maximization in social networks using viral marketing in the presence of other advertising channels.

\vspace{-1mm}
\section{The Proposed Framework
\vspace{-1mm}
}
\label{sec:model}
\noindent
We propose a framework to facilitate  study of different marketing aspects using a single model,
capturing several  factors:
% that are not considered in most of the information diffusion models
%\textbf{\color{red}{Give references.}}
\begin{enumerate}[leftmargin=*]
\item 
%{\em Multiple channels}: 
Companies market their products 
%to the nodes 
using multiple channels;
\item 
%{\em Clash of contagions}:
Diffusions of different products are mutually dependent;
% The influence of one product on a node may affect the influence of another product on it.
\item 
%{\em Personal aggregation}: 
Each node aggregates the mass media advertisements, recommendations, and neighbors' purchasing decisions.
%\item 
%%{\em Temporal changes in value}: 
%The enthusiasm of nodes to advertise a purchased product.
\end{enumerate}

We first describe  LT model, followed by our competitive multi-feature generalization, and then  integration of other channels into this generalized model.
%The notation is presented in 
Table \ref{tab:notation} presents notation.

%\textbf{\color{red}{
%First viral marketing using LT generalization, and then other channels integration
%}}

%As stated earlier, our model is a generalization of the LT model.
In LT model, every directed edge $(u,v)$ has weight $b_{uv} \geq 0$, which is the degree of influence that node $u$ has on node $v$, and every node $v$ has an influence threshold $\chi_v$. The weights $b_{uv}$ are such that $\sum_u b_{uv} \leq 1$. Owing to thresholds being private information to nodes, they are assumed to be chosen uniformly at random from $[0,1]$. The diffusion process starts at time step 0 with the initially activated set of seed nodes, and proceeds in discrete time steps. In each time step, a node gets influenced if and only if the sum of influence degrees of the edges incoming from its already influenced neighbors 
%(irrespective of the time of activation of the neighbors) 
crosses its influence threshold, that is, 
$
\sum_u b_{uv} \geq \chi_v
$.
%
%Nodes, once influenced, remain influenced for the rest of the diffusion process. 
% In any given time step, the recently activated nodes along with previously activated ones contribute to the diffusion process. 
The  process stops when no further nodes can be influenced.
%it is not possible to activate or influence any further nodes.
%
Formally, let $u\in \mathcal{N}(v)$ if and only if $b_{uv} \neq 0$. 
Let $\mathcal{B}(t)$ be the set of  nodes influenced by time $t$.
Then

\begin{table}[t!]
\vspace{-1.6mm}
\centering
\caption{Notation}
\label{tab:notation}
\vspace{-2mm}
\begin{tabular}{|p{.7cm}| p{7cm}|}
\hline 
%\hline
%\T \B
%$n$ & number of nodes in the network \\ \hline
%\T \B
%$v$ & a typical node in the network \\ \hline
\T \B
$b_{uv}$ & influence weight of node $u$ on node $v$ \\ \hline
\T \B
$\mathcal{N}(v)$ & set of influencing neighbors of $v$ \\ \hline
\T \B
$h_{uv}$ & similarity between nodes $u$ and $v$ \\ \hline
\T \B
$\chi_v$ & threshold of node $v$ \\ \hline
%\T \B
%$p$ & a typical product \\ \hline
%\T \B
%$f$ & number of features in a product \\ \hline
%\T \B
%$\delta^{p}$ & enthusiasm parameter for product $p$ \\ \hline
\T \B
$\gamma^p$ & total budget for the marketing of product $p$ \\ \hline
\T \B
$k^p$ & budget for seed nodes for viral marketing of product $p$ \\ \hline
\T \B
$\beta_t^p$ & mass media advertising weight of product $p$ in time step $t$ \\ \hline
\T \B
$\alpha^p$ & social advertising weight of product $p$ \\ \hline
\T \B
$\mathcal{A}_v$ & final aggregate preference of node $v$ \\ \hline
\T \B
$\mathcal{P}_{v}$ & product bought by node $v$ \\ \hline
\T \B
$\mathcal{B}(t)$ & set of nodes influenced by time $t$ \\ \hline
%\hline
\end{tabular}
\vspace{-4mm}
\end{table}

\begin{small}
\vspace{-3mm}
\begin{gather}
%\begin{split}
\hspace{-3mm}
v\in \mathcal{B}(t)\backslash \mathcal{B}(t-1)  \text{\;\;iff\;}   
    \sum_{\substack{u\in \mathcal{N}(v) \\ u\in \mathcal{B}(t-2)}} \!\!\!\!\!   b_{uv} < \chi_v \text{\;\;and} \sum_{\substack{u\in \mathcal{N}(v) \\ u\in \mathcal{B}(t-1)}} \!\!\!\!\!  b_{uv} \geq \chi_v
%\sum_{\substack{j\in \mathcal{N}(i) \\ j\in \mathcal{B}(t-1)}} b_{ij} \geq \mathcal{T}^{(i)} \text{\;and\;} \sum_{\substack{j\in \mathcal{N}(i) \\ j\in \mathcal{B}(t-2)}} b_{ij} < \mathcal{T}^{(i)}
%\end{split}
\label{eqn:LT}
\end{gather}
\vspace{-3mm}
\end{small}

\vspace{-1mm}
\subsection{Competitive Multi-feature Generalization of  LT Model
\vspace{-1mm}
}

%\textbf{\color{red}{Example of reals in model.}}

%\textbf{\color{red}{Example of features in phone}}

Products these days, be they toothpastes or mobile phones, come with several features with different emphases on different features.
%; we will assume these features to be independent of each other. 
Let such an emphasis be quantified by a real number between 0 and 1. 
That is, a product can be represented by a vector 
%consisting of elements which are 
of mutually 
independent features $p=(p_1,\cdots,p_f)$, where $p_i \in [0,1]$.
%
% or a vector itself, that is, $p_x $ or $p_x=<p_{x1},\cdots,p_{xm}>$.
Let the features of each product be suitably scaled such that $||p|| = 1$. Note that such scaling may not be feasible when there is a product $p$ which offers strictly better features than product $q$ ($\forall i:p_i>q_i$); so let the products be such that one of the features corresponds to `null'. So   $p$ would have a lower null component as compared to   $q$, thus making the scaling feasible (a higher null component would imply that the product has a poorer feature set).
%We do this by scaling all the product vectors by $\frac{1}{||p^{(\text{max})}||}$ where $||p^{(\text{max})}|| \geq ||p^{(z)}|| \; \forall z$, and then adding a ``Nothing'' dimension such that the products with the maximum $||p||$ have the ``Nothing'' component as $0$.

\begin{comment} %2nd
One can argue that a node should have different thresholds for different features; but as thresholds are generally assumed to be private information, it would only add to the uncertainty in the dynamics of diffusion. So we consider a common threshold for all the features put together, and not individually.
\end{comment} %2nd

%It is to be noted that even if products $p$ and $q$ are mutually orthogonally aligned, $p$'s influence may actually affect a node's decision regarding purchasing $q$ (since the threshold is an $f$-dimensional sphere and not cube). 

Our model for a node getting influenced is analogous to that used  in classical mechanics to study the initial motion of a body placed on a rough horizontal surface, as a result of several forces acting on it. In our context, a node is analogous to the body, and its threshold is analogous to static frictional force stopping it from moving. Such a force is equal to $\mu_s mg$, where $m$ is  mass of the body, $g$ is acceleration due to gravity, and $\mu_s$ is the coefficient of static friction between the body and  surface.
For $\mu_s g = 1$ unit for all nodes, the frictional force and analogously, the threshold  equals mass, which is chosen uniformly at random from $[0,1]$ (as assumed in the LT model). 

A node $v$ gets  influenced in time step $t$ when the net force on it crosses its threshold value $\chi_v$;
let the net force correspond to  aggregate vector (say $\mathcal{A}_v$).
Let $d(\mathcal{A}_v,p)$ be the angular distance between $\mathcal{A}_v$ and the product vector $p$. Since $||p||=1$,

\begin{figure}[t]
\vspace{-1.6mm}
\centering
\includegraphics[scale=.32]{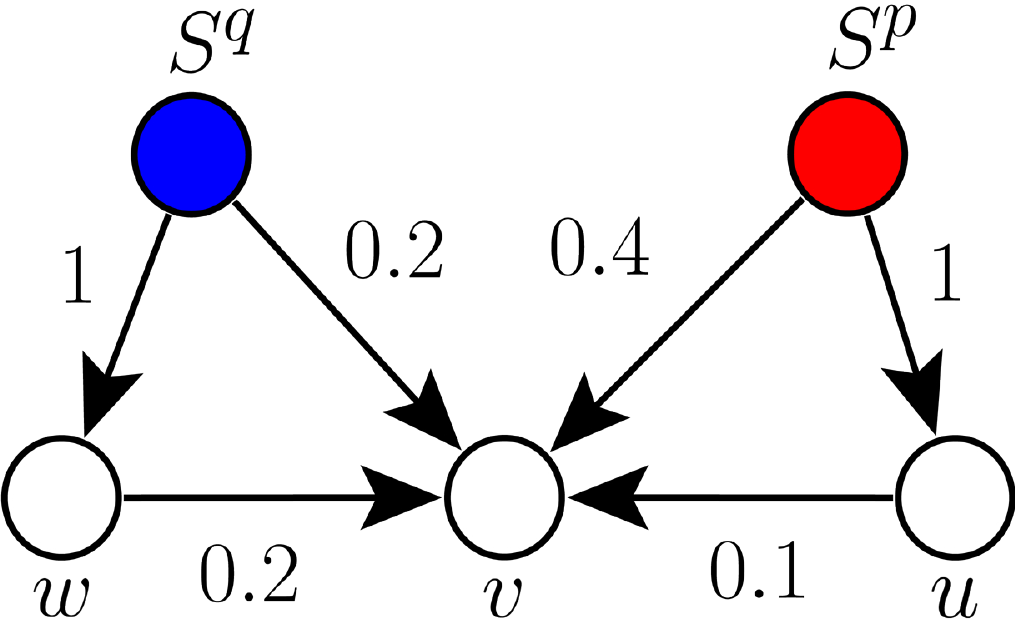}
\\
\includegraphics[scale=.55]{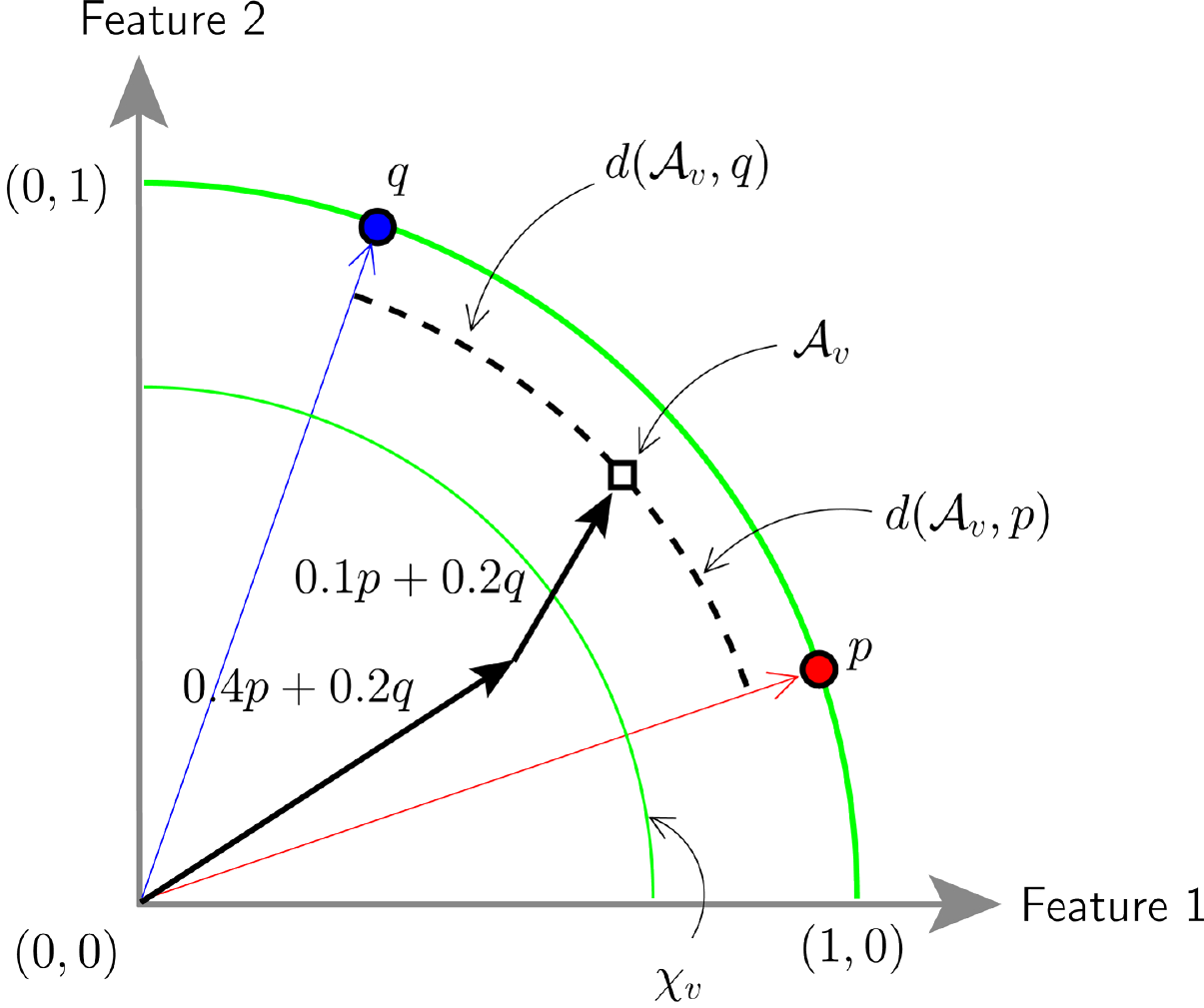}
\vspace{-2mm}
\caption{Geometric interpretation of the proposed model
%\textbf{\color{red}{label nodes 
%and change edge weights to account for delta
%}}
}
\label{fig:geometric_rep}
\vspace{-5mm}
\end{figure}

\begin{small}
\vspace{-2mm}
\begin{displaymath} 
d(\mathcal{A}_v,p) 
%= \arccos \left( \frac{\mathcal{A}_v \cdot p}{||\mathcal{A}_v|| ||p||} \right)
= \arccos \left( \frac{\mathcal{A}_v \cdot p}{||\mathcal{A}_v|| } \right)
%\;\;\;(\because ||p||=1)
\end{displaymath}
\vspace{-3mm}
\end{small}

A node buys a product whose angular distance from its aggregate vector is the least
(it can be easily shown that such a product would have the least Euclidean distance as well).
%note that this product is equivalent to buying a product whose Euclidean distance )
If there exist multiple such products, one of them is chosen uniformly at random. 

%2nd

%2nd

%Let $b_{uv}$ be the weight of influence of node $u$ on node $v$. 
%Let $u\in \mathcal{N}(v)$ if and only if $b_{uv} \neq 0$. 
%Let $\mathcal{B}(t)$ be the set of active nodes as observed at a given time $t$.
%Then

Hence the competitive multi-feature version of (\ref{eqn:LT}) is

\begin{small}
\vspace{-3mm}
\begin{gather*}
v\in \mathcal{B}(t)\backslash \mathcal{B}(t\!-\!1) \;,\; \mathcal{A}_v \!=\!\!\! \sum_{u\in \mathcal{N}(v)} \!\!\! {b_{uv}\mathcal{P}_u}
\;,\;
\mathcal{P}_v = \argmin_p d(\mathcal{A}_v,p)
%\\ 
%\text{iff\;\;\;}   
% \norm{ \sum_{\substack{u\in \mathcal{N}(v) \\ u\in \mathcal{B}(t-2)}} \!\!\!\!\!   b_{uv}\mathcal{P}_u} < \chi_v
%  \text{\;\;\;and\;\;\;}
% \norm{ \sum_{\substack{u\in \mathcal{N}(v) \\ u\in \mathcal{B}(t-1)}} \!\!\!\!\!  b_{uv}\mathcal{P}_u} \geq \chi_v
\end{gather*}
\vspace{-5mm}
\begin{gather*}
%v\in \mathcal{B}(t)\backslash \mathcal{B}(t\!-\!1) \;,\; \mathcal{A}_v \!=\!\!\! \sum_{u\in \mathcal{N}(v)} \!\!\! {b_{uv}\mathcal{P}_u}
%\;,\;
%\mathcal{P}_v = \argmin_p d(\mathcal{A}_v,p)
%\\ 
\text{iff\;\;\;}   
 \norm{ \sum_{\substack{u\in \mathcal{N}(v) \\ u\in \mathcal{B}(t-2)}} \!\!\!\!\!   b_{uv}\mathcal{P}_u} < \chi_v
  \text{\;\;\;and\;\;\;}
 \norm{ \sum_{\substack{u\in \mathcal{N}(v) \\ u\in \mathcal{B}(t-1)}} \!\!\!\!\!  b_{uv}\mathcal{P}_u} \geq \chi_v
\end{gather*}
\end{small}

%\textbf{\color{red}{Use delta for example}}

A geometric interpretation of the proposed model is presented 
%We now explain our generalization of the LT model with an example 
in Figure \ref{fig:geometric_rep}.
Consider 2 competing products having 2 features, say $p=(p_1,p_2),q=(q_1,q_2)$. 
%Let $\delta^p=\delta^q=1$.
In time step 0, $S_p$ and $S_q$ are selected for seeding by products $p$ and $q$, respectively.
In time step 1, node $v$ aggregates the purchasing decisions of its  neighbors $S^p$ and $S^q$, hence obtaining the aggregate vector $0.4p+0.2q$.
Say $\sqrt{(0.4p_1+0.2q_1)^2 + (0.4p_2+0.2q_2)^2} < \chi_v$, so $v$ is not influenced yet.
However, $u$ and $w$ purchase products $p$ and $q$ respectively (since the influence weights from $S^p$ to $u$ and $S^q$ to $w$ are 1). Hence in time step 2, node $v$ aggregates the purchasing decisions of $u$ and $w$, hence obtaining the aggregate vector $\mathcal{A}_v = (0.4p+0.2q)+(0.1p+0.2q)$.
Say $||\mathcal{A}_v||= \sqrt{(0.5p_1+0.4q_1)^2 + (0.5p_2+0.4q_2)^2} \geq \chi_v$, so $v$ is now influenced and it purchases product $p$ if $d(\mathcal{A}_v,p)<d(\mathcal{A}_v,q)$, $q$ if $d(\mathcal{A}_v,p)>d(\mathcal{A}_v,q)$, else it chooses randomly.
% if $d(\mathcal{A}_v,p)=d(\mathcal{A}_v,q)$.

%\subsection{Notes}
%Let all the products be normalized such that $||p^{(z)}|| = 1 \; \forall z$. We do this by scaling all the product vectors by $\frac{1}{||p^{(\text{max})}||}$ where $||p^{(\text{max})}|| \geq ||p^{(z)}|| \; \forall z$, and then adding a ``Nothing'' dimension such that the products with the maximum $||p||$ have the ``Nothing'' component as $0$.
%
%If a node does not get informed about a product through any channel at all,  that product becomes invisible to the node in the product space, and so even if the aggregate preference is closest to that product, the node opts for another product that is the closest among all the products that are visible to it.
%
%maximize $\nu(S,\alpha,\beta)$ \\
%such that \\
%$|S| = k$ \\
%$\alpha + k + \beta \leq \gamma$

\vspace{-1mm}
\subsection{Properties of the Generalized LT Model
\vspace{-1mm}
}

The standard LT model is a special case of the proposed model, where there is a single product with one feature, i.e., $p=(1)$.
% and $\delta^p=1$.
 As the problem of influence maximization in the standard LT model is NP-hard, we 
%get the following result.
have that 
the problem of influence maximization in the proposed model is also NP-hard.

%\begin{proposition}
%\label{prop:nphard}
%The problem of influence maximization in the proposed model is NP-hard.
%\end{proposition}

%\begin{displaymath}
%\frac{ \sum_{u \in A_{t} \backslash A_{t-1}} b_{vu} \delta ^t }{ 1 - \sum_{s=0}^{t-1} \sum_{u \in A_{s} \backslash A_{s-1}} b_{vu} \delta ^s }
%\end{displaymath}

\begin{wrapfigure}{l}{27mm}
\vspace{-.3cm}
%\hspace{-10mm}
  %\begin{center}
\includegraphics[scale=.31]{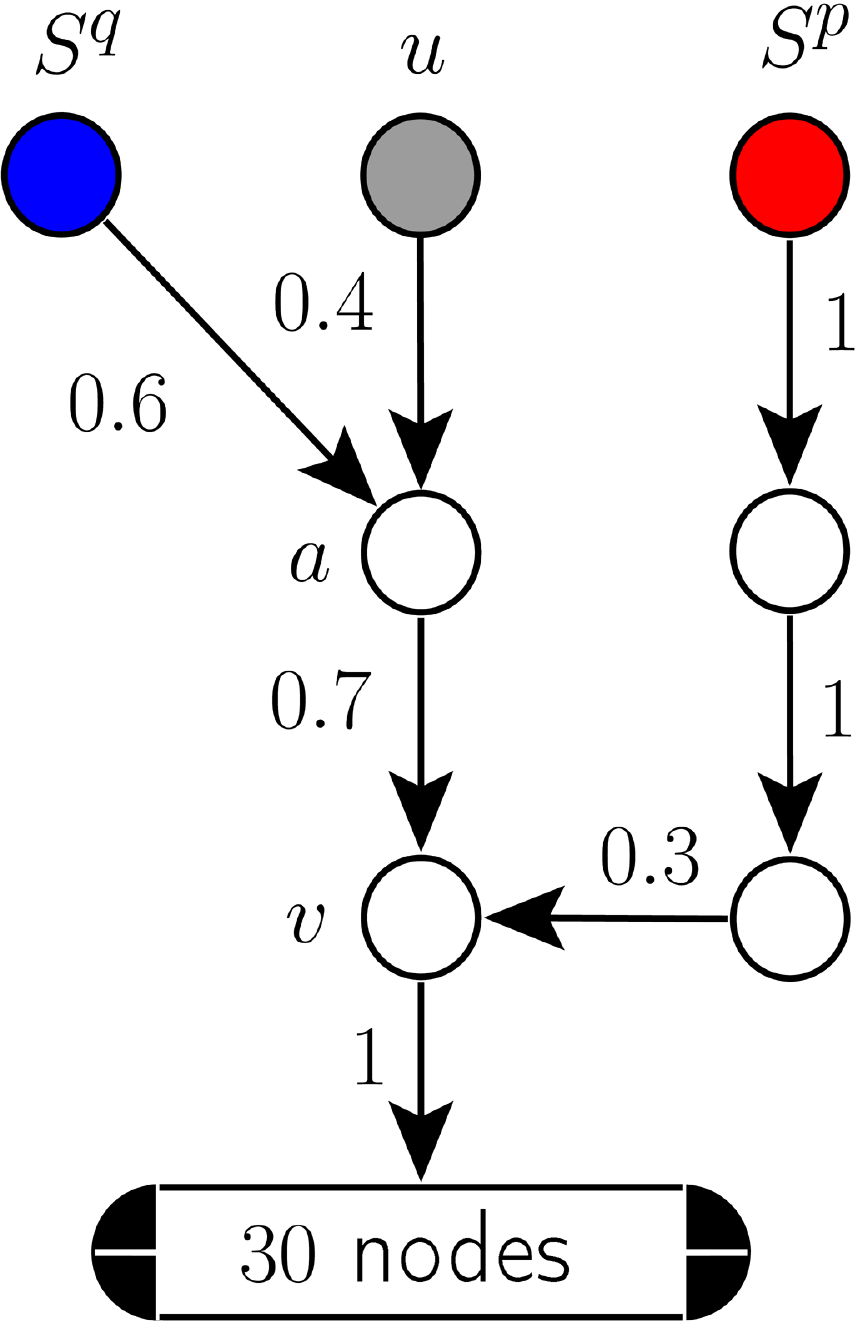}
  %\end{center}  
  \vspace{-.6cm}
\caption{
%\hspace{-10mm}
Example 
%to show non-monotonicity of the influence maximization objective function under the proposed model
} 
\vspace{-.3cm}
\label{fig:countereg_monotone}
\end{wrapfigure}

%Moreover, it can be seen with a small example shown in Figure \ref{fig:countereg_monotone} that the influence maximization objective function $\sigma(\cdot)$ under the proposed model is not even monotone.
We now explain the multi-feature (vector-based) model with an illustrative example, which will also throw light on the properties of the objective function under the proposed model.
%, that is,
%\begin{displaymath}
%S^p \subset T^p \centernot\implies \sigma(S^p) \leq \sigma(T^p)
%\end{displaymath}
%
Recollect that the threshold for any node is chosen uniformly at random from $[0,1]$.
Let $\mathbb{P}_v^p (S^p,S^q)$ be the probability that node $v$ gets influenced by product $p$ when $S^p$ and $S^q$ are selected for seeding by $p$ and $q$ respectively.
In Figure \ref{fig:countereg_monotone}, let the two products be $p=(1,0)$ and $q=(0,1)$.
%, with $\delta^p=\delta^q=1$. 
%A node would buy a product, whose angular distance from its aggregate preference immediately after crossing its threshold, is the least. 
%
%
%\textbf{\color{red}{No greedy. Other algorithm or Shapley value.}}
\hspace{-10mm}
Let $\sigma^p(S^p,S^q)$ be the expected number of nodes influenced by $p$ when $S^p$ and $S^q$ are selected for seeding by $p$ and $q$ respectively. 
That is, $\sigma^p(S^p,S^q) = \sum_i \mathbb{P}_i^p (S^p,S^q)$.
%the seed set for $p$ be $S^p$ as labelled and that for $q$ be $S^q$. 
With  diffusion starting from $S^p$ and $S^q$ simultaneously,
we have
$\mathbb{P}_a^p (S^p,S^q) = 0$ and  
$\mathbb{P}_a^q (S^p,S^q) = 0.60$.
Note that if $a$ is influenced by $q$, then it influences  $v$ with probability 0.70 even before the influence of $p$ reaches it, starting from $S^p$; now even if influence of $p$ reaches it, it is impossible for its aggregate preference to be closer to $p$ than to $q$. 
So node $v$ can get influenced by $p$ only if $a$ is not influenced by $q$. 
So $\mathbb{P}_v^p (S^p,S^q) = 0.3(1-\mathbb{P}_a^q (S^p,S^q)) = 0.12$. So all 30 nodes which have  $v$ as sure influencer get influenced by $p$ with probability 0.12.
Hence $\sigma^p (S^p,S^q) = 1+2+0.12(1+30) = 6.72$.

Now if the seed set for $p$ is $T^p = S^p \cup u$, $\mathbb{P}_a^p (T^p , S^q) = 0$   due to an incoming edge of 0.6 from $S^q$ (it is  impossible for the aggregate preference of node $a$ to be closer to $p$ than to $q$). However, 
$\mathbb{P}_a^q (T^p , S^q) = \sqrt{0.6^2+0.4^2} \approx 0.72$.
From the argument similar as above,  $\mathbb{P}_v^p (T^p , S^q) = 0.3(1-\mathbb{P}_a^q (T^p , S^q)) < 0.084$.
So the  30 nodes 
%which have  $v$ as a sure influencer 
get influenced by $p$ with probability less than 0.084.
Hence $\sigma^p (T^p ,S^q) < 2+2+0.084(1+30) < 6.61$.
That is, $\sigma^p (T^p ,S^q) < \sigma^p (S^p,S^q)$.

Thus while the objective function in the standard LT model follows monotone increasing property,
 adding a node to a set in the generalized model could decrease its value;
this proves {non-monotonicity}.
It can also be shown using counterexamples that $\sigma^p(\cdot)$ is neither submodular nor supermodular.
%It can also be shown using examples that, it is possible to find a tuple $(S^p,T^p,u)$ satisfying $S^p \subset T^p \subseteq N$ and $u \in N\setminus T^p$, such that
%$\sigma^p(T^p \cup u) - \sigma^p(T^p) > \sigma^p(S^p \cup u) - \sigma^p(S^p)$, while we can find another tuple that reverses the inequality.

%Hence we have that
%get the following result.
%
%\begin{proposition}
%$\sigma^p(\cdot)$ is not monotone and neither submodular nor supermodular.
%\end{proposition}
%
\begin{comment} %2nd
However, it was observed using simulations that the objective function $\sigma^p(\cdot)$ was close to being monotone increasing, that is, the monotonicity property was satisfied in most instances, while it was far from being either submodular or supermodular.

The above example also shows that even if products $p$ and $q$ are mutually orthogonally aligned in space, $p$'s influence may affect a node's decision regarding purchasing $q$ (since the threshold is an $f$-dimensional sphere and not cube). 
\end{comment} %2nd

\begin{figure}[t]
\vspace{-1.2mm}
\centering
\includegraphics[scale=.42]{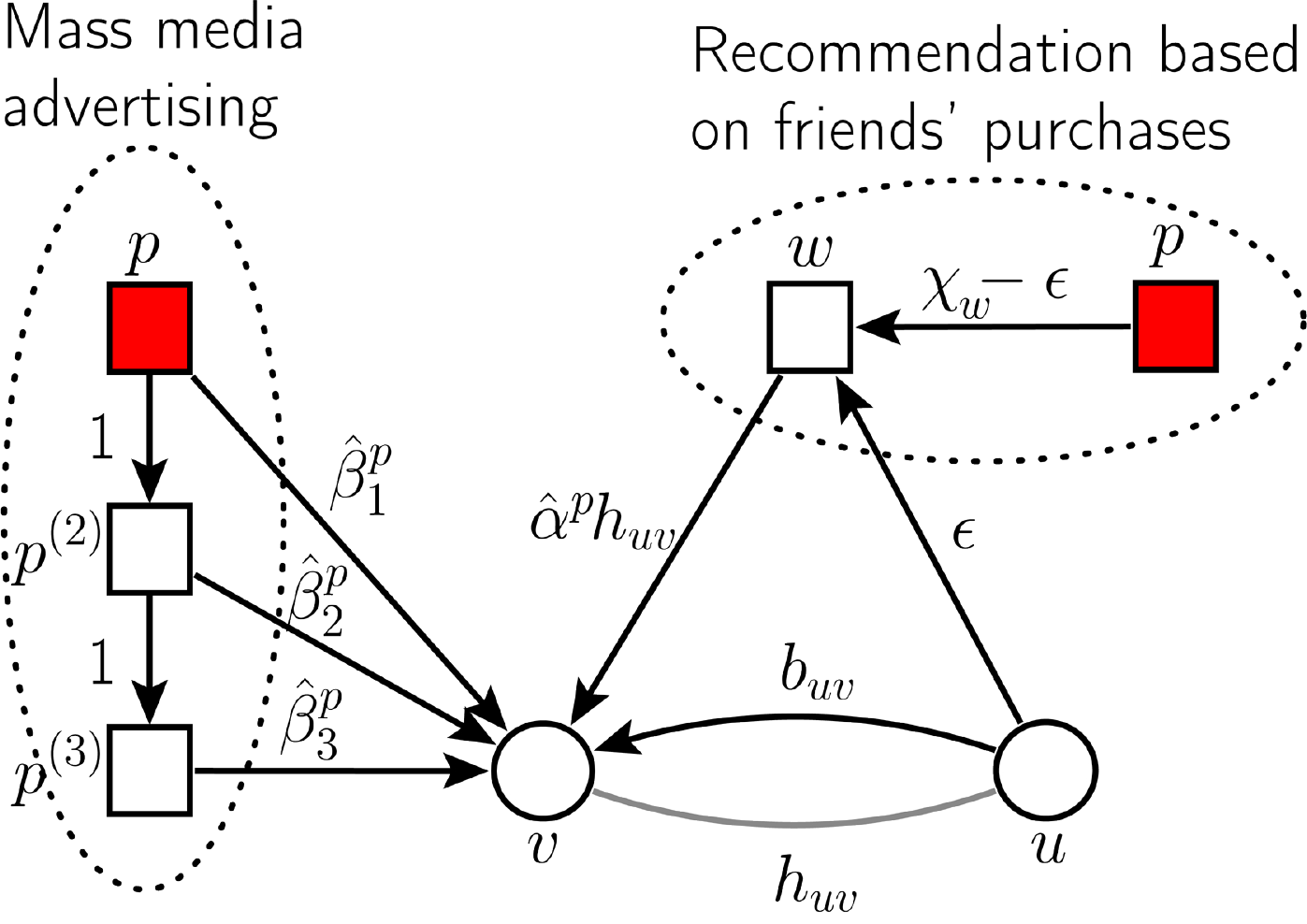}
\vspace{-2mm}
\caption{Integration of mass media and social advertisements into the  network 
%\textbf{\color{red}{change u,v and put beta/alpha hat}}
}
\label{fig:integration}
\vspace{-4mm}
\end{figure}

\vspace{-1mm}
\subsection{Integrating Mass Media \& Social Advertising into  Network\vspace{-1mm}
}

Let  $\beta_t^p$ be the investment for mass media advertising of product $p$ in time step $t$ and $\beta^p$ be the total investment over  time $T$, that is, $\beta^p = \sum_{t=1}^T \beta_t^p$.
For social advertising, we consider that a company would recommend or advertise product $p$ to a node when any of its friends $u$ has bought the product. Let $\alpha^p$ be the effort invested in social advertising. 
Let $h_{uv}$ (or $h_{vu}$) be the parameter that quantifies the similarity between nodes $u$ and $v$. 
%For the above reasons, we assume that
So $\alpha^p h_{vu}$ could be viewed as the influence of such a  recommendation on  $v$ owing to the purchase of product $p$ by  $u$. 
%We also assume the total influence of recommendation on node $v$ to be a sum of such indirect recommendations from all of its neighbors ($u$'s) who have purchased the product $p$.
%
Since the total influence weight allotted by node $v$ for viral marketing is $\sum_{u \in \mathcal{N}(v)} b_{uv}$, the total weight that it can allot for other channels is $1-\sum_{u \in \mathcal{N}(v)} b_{uv}$.
Hence the weights allotted for other channels ($({\beta}_t^p)_{t=1}^T$ and ${\alpha}^p$) would be scaled accordingly to obtain the values of $(\hat{\beta}_t^p)_{t=1}^T$ and $\hat{\alpha}^p$ specific to node $v$. A simple scaling rule is:

\begin{small}
\vspace{-2mm}
\begin{gather*}
\frac{\hat{\beta}_t^p}{{\beta}_t^p} = \frac{\hat{\alpha}^p}{\alpha^p} =  \frac{1-\sum_{u \in \mathcal{N}(v)} b_{uv}}{ \sum_p \left( \sum_{u \in \mathcal{N}(v)} \alpha^p h_{uv} + \sum_{t=1}^T \beta_t^p \right)} 
\end{gather*}
\vspace{-2mm}
\end{small}

In order to integrate mass media and social advertisements into the network, we add pseudonodes and pseudoedges corresponding to them, as illustrated in Figure \ref{fig:integration}.
Pseudonode $p$ corresponds to the product company itself (the figure shows two separate copies of pseudonode $p$ for the two channels for better visualization; they are the same pseudonode).
Pseudonode $p$ and all seed nodes selected for viral marketing, are influenced in time step 0.

For integrating mass media advertisement,
we create a set of pseudonodes $\{p^{(t)}\}_{t=1}^T$ (where $p^{(1)}$ corresponds to pseudonode $p$), and pseudoedges $\{(p^{(t-1)},p^{(t)})\}_{t=2}^T$
of weight 1.
%such that $b_{pp'}=1,b_{p'p''}=1,$ and so on.
Hence  $p^{(t)}$ gets influenced with probability 1 in time step $t-1$ (see Figure \ref{fig:integration}).
% and so on, respectively.
We further create pseudoedges $\{(p^{(t)},v)\}_{t=1}^T$ for node $v$ such that $b_{p^{(t)} v} = \hat{\beta}_t^p$.
%
%We create pseudoedges with weights $\hat{\beta}_1^p,\hat{\beta}_2^p,\hat{\beta}_3^p,$ and so on from pseudonodes $p,p',p'',$ and so on, respectively to node $v$.
%That is, $b_{pv}=\hat{\beta}_1^p, b_{p'v}=\hat{\beta}_2^p, b_{p''v}=\hat{\beta}_3^p$, and so on.
%
%
%\subsection{Some Results}
%\label{sec:theory}
%
%\begin{proposition}
%\label{prop:mass_ad}
%The mass media advertisements can be integrated in the social network using Construction~\ref{cns:model}.
%\end{proposition}
%\begin{proof}
Since $p^{(t)}$ gets influenced in time  \mbox{$t-1$},
%It can thus be  seen that   
node $v$ receives  influence  of $\hat{\beta}_t^p$ from pseudonode $p^{(t)}$ in time step $t$; this is equivalent to  mass media advertisement.
%Thus the mass media advertisement of product $p$ is integrated into the network.
%\end{proof}

%\vspace{-1mm}
%\subsection{Integrating Social Advertising into the Network
%\vspace{-1mm}
%}

For integrating social advertisement, corresponding to edge $(u,v)$, we create an intermediary pseudonode $w$ with a fixed threshold $\chi_w>0$, and pseudoedges such that $b_{uw}=\epsilon \in (0,\chi_w), b_{pw}=\chi_w - \epsilon, b_{wv} = \hat{\alpha}^p h_{uv}$ (see Figure \ref{fig:integration}).
%
%\begin{construction}
%\label{cns:model}
%All advertising nodes are activated at time step 0.
%Seed nodes are selected at time step 1.
%Separate intermediary as well as advertising nodes required for social advertising integration into the social network. 
%It is necessary that $\epsilon \in (0,\chi)$.
%\end{construction}
%\textbf{\color{red}{Complete it}}
%
%\begin{proposition}
%\label{prop:social_ad}
%The recommendations can be integrated in the social network using Construction~\ref{cns:model}.
%\end{proposition}
%\begin{proof}
%Let $\chi$ be the threshold of the intermediary pseudo-node and let $\epsilon \in (0,\chi)$. 
Now if the reference friend $u$ is influenced by some product $q$, where the angle between products $p$ and $q$ be $\theta$, the intermediary pseudonode $w$ gets influenced if and only if 
$|| (\chi_w - \epsilon)p + \epsilon q || \geq \chi_w$. Since $||p|| = ||q|| = 1$, this is equivalent to

~
\begin{small}
\vspace{-6mm}
\begin{align*}
%&  || (\chi - \epsilon)p + \epsilon q || \geq \chi \\
% &\Leftrightarrow 
%(\chi - \epsilon)^2 ||p||^2 + \epsilon ^2 ||q||^2 + 2\epsilon (\chi - \epsilon) ||p|| ||q|| \cos \theta \geq \chi ^2 \\
 %&\Leftrightarrow 
%& \;\;\;\;\; 
&(\chi_w - \epsilon)^2 + \epsilon ^2 + 2\epsilon (\chi_w - \epsilon) \cos \theta \geq \chi_w ^2 \\
 \Longleftrightarrow \;&
2 \epsilon (\chi_w - \epsilon)(\cos \theta - 1) \geq 0 \\
%\;\;\; (\because ||p|| = ||q|| = 1) \\
\Longleftrightarrow \;&
\theta = 0 \;\;\; (\because \epsilon < \chi_w ) \\
\Longleftrightarrow \;&
q = p \;\;\; (\because ||p||=||q|| )
\end{align*}
\vspace{-4.2mm}
\end{small}

%Hence node $v$ is recommended product $p$ 
So $w$ gets influenced if and only if  $u$ buys product $p$, after which  $v$ is recommended to buy $p$ with influence weight $\hat{\alpha}^p h_{uv}$.
Also note the time lapse of one step between the reference friend $u$ buying the product and the target node $v$ receiving the recommendation. Hence the latency in recommendation using social advertising is implicitly accounted for.
%\end{proof}

%\textbf{\color{red}{
%buv scaled down since original buv was only for viral marketing, and new one for all channels
%\\
%sum of all incoming edges and pseudoedges at most 1
%}}

\begin{comment} %2nd
It is to be noted that the original values of $b_{uv}$ assume that the influence on a node $v$ is entirely due to viral marketing, and do not account for influence through other channels.
With other influencing channels, the influence weight allotted by $v$ to decision of $u$ would reduce.
Since now the sum of incoming edges and pseudoedges is
$
\sum_{u \in \mathcal{N}(v)} b_{uv}+ \sum_p \left( \sum_{u \in \mathcal{N}(v)} \alpha^p h_{uv} + \sum_{t=1}^T \beta_t^p \right)
$,
the scaled down influence weights are

\begin{small}
\vspace{-2mm}
\begin{gather*}
\hat{b}_{uv} = b_{uv}\left( \frac{\sum_{u \in \mathcal{N}(v)} b_{uv}}{\sum_{u \in \mathcal{N}(v)} b_{uv}+ \sum_p \left( \sum_{u \in \mathcal{N}(v)} \alpha^p h_{uv} + \sum_{t=1}^T \beta_t^p \right)} \right)
\end{gather*}
\end{small}
\end{comment} %2nd

%\textbf{\color{red}{or fit inside 1-sum buv}}

\vspace{-1mm}
\section{The Underlying Problem 
\vspace{-1mm}
}
\label{sec:problem}

%2nd
The fundamental problem here is to distribute the total available budget among the three marketing channels under study. 
%We use the term `effort' very loosely here. What we actually mean is that, more the effort put into a particular type of marketing, (a)  more is its effectiveness and (b) more is the accompanying cost.
%
%Let $\beta_t^p$ be the effort spent for mass media advertising of product $p$ at time step $t$.
%So $\beta^p = \sum_{t=1}^T \beta_t^p$.
%\begin{displaymath}
%\beta^p = \sum_{t=1}^T \beta_t^p
%\end{displaymath}
%
%Let $\alpha^p$ be the effort spent for additional social advertising.
Let $k^p$ be the number of free samples of product $p$ that the company would be willing to distribute. 
Let $S^p$ be the corresponding set of nodes in the social network to whom free samples would be provided ($|S^p|=k^p$).
%Also, given a particular value of $k^p$, assume for now that, the company can uniquely determine $S^p$, the set of nodes in the social network to whom free samples would be provided.
%
%
%
%
%\textbf{\color{red}{
%difference between sigma and nu
%\\
%weighted sum?
%}}
%
%
%
Let $c_p(\cdot)$ be the cost function for allotting effort of activating set $S^p$ to trigger viral marketing, $\alpha^p$ for social advertising, and $(\beta_t^{q})_{t=1}^T$ corresponding to each step of mass media advertising.
In general, $c_p(\cdot)$ would be a weighted sum of these parameters since the costs for adjusting parameters corresponding to different channels would be different.
Let $\gamma^p$ be the total budget for marketing of product $p$. 
Let $\nu^p(\cdot)$ be the expected number of nodes (excluding pseudonodes) influenced by product $p$, accounting for the marketing strategies of $p$ and its competitors. 
Hence the optimization problem for the marketing of product $p$ is

\begin{small}
\vspace{-3mm}
\begin{equation}
\label{eqn:optimization_problem}
\begin{split}
\text{Find } S^p,\alpha^p,(\beta_t^p)_{t=1}^T \text{ to maximize }\\
 \nu^p(S^p,\alpha^p,(\beta_t^p)_{t=1}^T,(S^{q},\alpha^{q},(\beta_t^{q})_{t=1}^T)_{q \neq p}) \\
\text{ such that }
c^p(S^p,\alpha^p,(\beta_t^p)_{t=1}^T )  \leq \gamma^p
\end{split}
\end{equation}
\vspace{-2mm}
\end{small}

%2nd

%\vspace{-1mm}
%\subsection{A General Approach
%\vspace{-1mm}
%}

%When we move away from standard simplistic models, we are less likely to obtain seed selection algorithms with approximation guarantees.
%The proposed framework also leads to the objective function losing nice properties such as monotonicity and submodularity. 
%We hence propose adapting a well-known algorithm for the purpose of such multi-parametric optimization which does not follow nice properties.

\begin{comment} %2nd
With $\sigma^p(\cdot)$ not having any particular property, it is not guaranteed that the greedy algorithm, which performs well for optimizing monotone submodular functions, will perform well for optimizing $\sigma^p(\cdot)$.
In what follows, let $\mathcal{T}$ be the time taken for computing the objective function value for a given set.
%
%We propose the usage of cross entropy method for the considered optimization problem.
\end{comment} %2nd

%\subsection{Candidate Algorithms}

%\subsubsection{Fully Adaptive Cross Entropy Method (FACE)}
%\label{sec:ce_method}

In the above optimization problem, we not only need to determine the optimal allocation among channels ($k^p,\beta^p,\alpha^p$), but the best $k^p$ nodes to trigger viral marketing $S^p$ such that $|S^p|=k^p$, and the optimal allocation over time for mass media advertising $(\beta_t^p)_{t=1}^T$ such that $\sum_{t=1}^T \beta_t^p = \beta^p$.
The problem hence demands a method for
%We hence propose adapting a known method for the purpose of such 
multi-parametric optimization.
% which does not follow nice properties such as monotonicity, submodularity, and supermodularity.

%\textbf{\color{red}{
%Furthermore, budget allocation into k,beta,alpha. k further find optimal set. beta further find optimal temporal allocation. FACE does simultaneous optimization
%}}

%This algorithm would not only find an optimal seed set, but also implicitly determine how to split the total budget among the channels.

Methods such as Fully Adaptive Cross Entropy  (FACE) 
%It has been shown that the cross entropy (CE) 
 provide a simple, efficient, and general approach for simultaneous optimization over several parameters~\cite{de2005tutorial}.
% where the problem is completely known and static, as well as noisy estimation problems where the unknown objective function needs to be estimated using techniques such as discrete event simulation~\cite{de2005tutorial}.
%It can be viewed as a generic and practical tool for solving NP-hard problems.
 %\cite{rubinstein1999cross}.
In our context, the FACE method involves an iterative procedure where each iteration consists of two steps, namely,
(a) generating data samples according to a specified distribution and
(b) updating the distribution based on the sampled data to produce better samples in the next iteration.
%
%
%\textbf{\color{red}{
Here, our sample is a vector consisting of 
whether a node should be included in $S^p$,
%a sampled 
%$k_1$ and $d$ and also a 
%candidate seed set $S^p$,
budget allotted for each time step of mass media advertising $(\beta_t^p)_{t=1}^T$, and budget allotted for social advertising $\alpha^p$;  each data sample satisfies the cost constraint $c^p(S^p,\alpha^p,(\beta_t^p)_{t=1}^T )  \leq \gamma^p$.
%}}
%
Initially, the data samples could be generated based on a random distribution.
The value of the objective function $\nu^p(\cdot)$ is computed for each data sample as per the proposed model using a sufficiently large number of Monte Carlo simulations.
The distribution is then updated by considering data samples which provide value of the objective function better than a certain percentile.
This iterative updating continues until convergence or for a fixed number of iterations.
The obtained terminal data sample acts as the best response allocation strategy for product $p$, in response to the strategies of  competitors.

% %For our simulations, 
%% We use an adaptive version of the CE method called the {\em fully adaptive cross entropy} (FACE) algorithm~\cite{de2005tutorial}.
% Its time complexity is $O(n\mathcal{T} \mathcal{I})$, where 
% $\mathcal{T}$ is the time taken for computing the objective function value for a given set and 
% $\mathcal{I}$ is the number of iterations before the algorithm terminates.
% %

%An immediate future work is to observe budget allocations 

\begin{comment} %2nd
For the detailed FACE algorithm and the terminology used, the reader is referred to \cite{de2005tutorial}.
We initialize the method with distribution $(\frac{\gamma}{n},\ldots,\frac{\gamma}{n})$, that is, each node has a probability of $\frac{\gamma}{n}$ of getting selected in any sample set in the first iteration.
%(where $\gamma$ is the budget which is $k,k_1,k_2$ for single phase diffusion, first phase, and second phase, respectively).
In any iteration, the number of samples (satisfying budget constraint) is bounded by
 $N_{\text{min}}=n$ and
$N_{\text{max}}=20n$, and the number of elite samples is
$N_{\text{elite}}=\lceil \frac{n}{4} \rceil$.
%$n$ and $2n$, and the number of elite samples is $N_{\text{elite}}=\lceil \frac{n}{4} \rceil$.
We use a weighted update rule for the distribution, where in any given iteration, the weight of any elite sample is proportional to its function value. %~\cite{de2005tutorial}.
The smoothing factor that we consider is $\alpha = 0.6$.
%$l = ceil(n/20)$,
%$d = ceil(n/20)$
%
\end{comment} %2nd

\vspace{-1.2mm}
\subsection*{A Note for Practical Implementation
\vspace{-1mm}
}

In order to implement the proposed framework in practice, a company would need to map its customers to the corresponding nodes in  social network. To create such a mapping, it would be useful to get the online social networking identity (say Facebook ID) of a customer as soon as it buys the product. 
This can be done using a product registration website (say for activating warranty) where a customer, when it buys the product, needs to login using a popular social networking website (such as Facebook), or needs to provide its email address which could be used to discover its online social networking identity.
Thus the time step when the node has bought the product, can also be obtained.
%Once a purchase is detected, the company can recommend the product to its friends.

\begin{comment} %2nd
\section{Discussion}
\label{sec:discuss}
The problem of competitive influence maximization for a company is a complex one. There are several resources available and the company has to invest appropriately in the appropriate resources at appropriate times. The companies need to constantly keep on revising their strategies based on the past outcomes, which in turn, depend on the strategies of the competitors as well as the behavior of the potential customers (in the sense that how they aggregate the information they get from various channels). The emerging possibility of viral marketing through online social networks has added a new dimension to the problem. Our objective in this work was to present a framework for studying optimal strategies 
%(selecting seed nodes) for triggering viral marketing, 
in the presence of several marketing channels, so as to maximize the spread of influence in the social network.
\end{comment} %2nd

\vspace{-1mm}
\section*{References
\vspace{-1.25mm}
}
%\newpage
%\newpage
%\newpage
\bibliographystyle{IEEEtran}
%\begin{small}
%\bibliographystyle{IEEEtran}
\begingroup
\renewcommand{\section}[2]{}%
\begin{small}
\bibliography{cidp_references} 
\end{small}
\endgroup 

%\vspace{-2mm}
%\begin{small}
%\bibliographystyle{IEEEtran}
%\bibliography{cidp_references}  
%\end{small}
% You must have a proper ".bib" file
%  and remember to run:
% latex bibtex latex latex
% to resolve all references
%
% ACM needs 'a single self-contained file'!
%
%APPENDICES are optional
%\balancecolumns
%\appendix

\end{document}